%% file: chapter.tex
\documentclass[12pt,reqno]{amsart}
\usepackage{graphicx}
\usepackage{amssymb,amscd,amsbsy}
\usepackage{amsthm}
\usepackage{pdfpages}

\usepackage{amsmath}
\usepackage{amsfonts}   
\usepackage{amssymb} 

\input{definitions}

\theoremstyle{remark}

\newtheorem*{rem*}{Remark}

\begin{document}

\newcommand{\vse}{\vspace{.2in}}
\numberwithin{equation}{section}

\title{On   Simulation of Various Effects in Consolidated Order Book. }
\author{A.O. Glekin, A.  Lykov and K.L. Vaninsky}
\begin{abstract}
This paper consists of two parts. The first part is devoted to empirical analysis  of consolidated order book (COB) for the index RTS futures. In the second part  we consider Poissonian  multi--agent model of the COB. By varying parameters of different groups of agents submitting 
orders to the book we are able to model various real life  phenomenas. In particular we model  the  spread, the  profile of the book and large price changes. Two different mechanisms of large  price changes are considered in detail. One is  the disbalance  of liquidity in the  COB  and another is the disbalance of sell and buy orders in the order flow.   
\end{abstract}
\maketitle

\tableofcontents 
\setcounter{section}{0}
\setcounter{equation}{0}

\section{Introduction. } 

Price  changes and its causes been  a classical topic of economics research for a long time  already. The answer  to the  traditional question "why prices change" in the theory of effective market is that the market absorbs a new information which forces market participants to reconsider the  price of securities, currencies, futures,  {\it etc}. 

At the level of micro--structure   investigation of a price change became possible only after historical data  about all orders and events in the exchange  became publicly available. For a novice in the field we  say  that the order matching mechanism of an   exchange is called consolidated order book (COB) or simply the  book. 
In this work we do not consider the causes that determine the rate of submitting   limit and market orders to the book by market participants. 
Instead the main emphases  is made on a study of various book statistics and a price changing mechanisms.    
In this work we assume that all rates are constant in time  {\it i.e.} we stay in the realm  of "zero" intelligence traders, \cite{GS}. 

It turns out that there are two  basic mechanisms  of the  price change. In one case it is the disbalance between demand through the flow of market orders 
and supply of limit orders in the book. Nevertheless,  the simple rule telling us what is happening  if the demand exceeds supply
not necessarily leads to a price change. Another cause  of the price change is  the disbalance  of liquidity in the order book. 
In reality two of these mechanisms contribute in a certain combination.

Our work consists of two parts. The first part  is devoted to   study  of empirical statistics of the book and the order flow for   futures on the  RTS index. 
Similar  investigation was previously performed on  stocks  traded on French and USA equity exchanges  \cite{BHS}  and  \cite{BMP}.
The Russian Trading System (RTS)  is a stock market established in 1995 in Moscow, consolidating various regional trading floors into one exchange. 
Originally RTS was modeled on NASDAQ's trading and settlement framework.   
The RTS Index, RTSI, the official Exchange indicator, first calculated on September 1, 1995, is similar  to the Dow Jones 
Index.  Nowadays the value of contracts traded  in RTS Index futures and options   exceeded tens of  billion dollars.  The number of open positions (open interest) exceeds  250,000 contracts. The excellent liquidity allows us to compute  various statistics  of the order book from historical data provided by RTS.

The second part of this work contains simulations performed with the use of the Poissonian  multi--agent model of the order book. 
For the first time Poissonian models were considered by J.D. Farmer {\it at all} in  \cite{SFGK, DFGIS, MF}.
We formulate our model using multi--agent framework \cite{CTPA2, SZSL}. 
We  presents results of numerical simulations  which are   similar to  real statistics of the  RTS index futures.  
We have to mention that  some of these  simulations already appeared  
in \cite{G}.  The limiting  case of the model corresponding to the book of density one was rigorously considered by us in \cite{MLV}  and \cite{MV}.

For convenience of the reader we start our presentation  with  detailed description of the order book.

\section{The Order Book. }

In this work we consider   an exchange  with    continuous double auction as the  order matching mechanism.  
Market participants  submit  to the exchange orders  of two types, namely limit orders and market orders.

The  limit orders are specified by  three parameters, the price level, the volume and the direction (buy or sell). 
The price is the worst price at which the order can be executed. 
The volume of an order is  a number of contracts which constitute the order.

The consolidated order book is shown in  Figure 1. The mid--column is  a price ladder for a security. 
Each limit order is placed in the order book at the  level  specified by its price.
The minimal price to sell is called an ask price and the maximal price to buy is called a bid price. 
The first column represents a  price level counted from the best ''bid'' price. 
The second  ''bid'' column represents a  total volume  of orders that can be bought at the specific price. 
On the right from the middle column the situation is identical  but reversed. The fourth column is the total volume of orders at the specific price level. The 
last  the fifth column is a price level counted from the best ''ask''.

\begin{figure}[htp] 
\centering{
\includegraphics[scale=0.82]{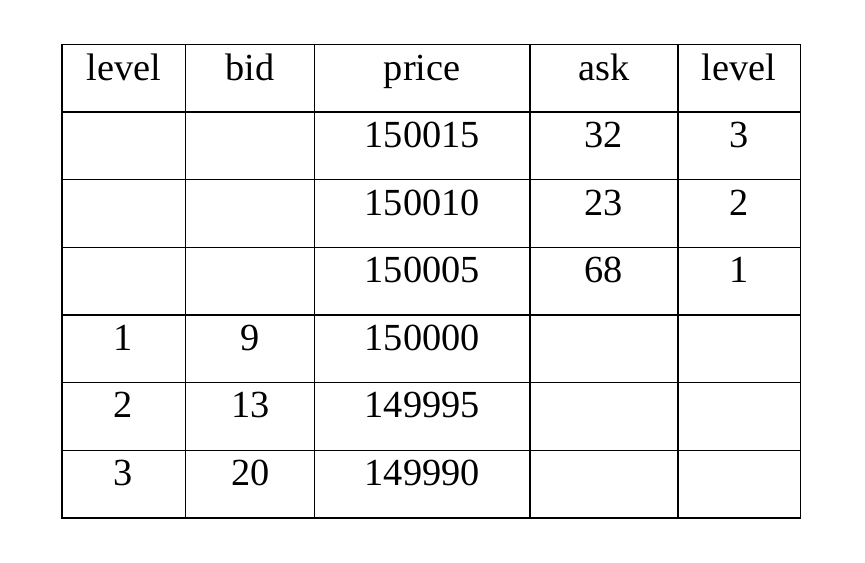}}
\caption{Consolidated Order Book}
\end{figure} 


\noi
Limit order stay in the book until they get executed or just canceled. Execution of  orders in each queue is determined  by the  FIFO rule {\it i.e.} 
First In,  First Out.

The price level for ''buy''  orders  is counted from the best price offer (ask price) at the moment 
$$ 
l(p)=\frac{p_{ask}-p}{s}, \qquad\qquad p<p_{ask}; 
$$
where $p_{ask}$ is the best price to buy  and $s$ is the size of the price ladder  step.  
Similarly the price level  for  ''sell'' orders is counted from the best price offer (bid price) 
$$
l(p)=\frac{p-p_{bid}}{s}, \qquad \qquad   p>p_{bid};
$$
where $p_{bid}$ is the best price to sell.

The state of the book is given by a vector $X=\{   x_i \}$, where  $|x_i|$  is  aggregated volume of  orders at level $i$.  The component $x_i$ is positive if 
these are buy orders and negative if these are sell orders. 
Note that 
$$
i=\frac{price}{s}, \quad\quad\quad i_{ask} =\frac{p_{ask}}{s}, \quad\quad\quad  i_{bid} =\frac{p_{bid}}{s}. 
$$
We define  instant liquidity to sell  as  
$$
s(l)=\sum_{i=i_{ask}}^{i_{ask}+l} X_{i}, \quad\quad s=s(\infty);
$$
and instant liquidity to buy as 
$$
d(l)=\sum_{i=i_{bid}}^{i_{bid}-l} X_{i},\quad\quad  d=d(\infty). 
$$


Market orders are the orders which have no specific price and the only volume is specified. Such orders are executed at the best available price  at the moment they are 
submitted. If, for example,  in the book shown in  Fig 1 submitted a market order to buy of the size 70, then the part of it (68 orders) is executed at the price 150005  and the remaining 2 orders are executed at the price 150010.


\section{The Empirical Statistics of the RTS Futures  Order Book.} 
\subsection{The rate of submitting or canceling orders.} 
The rate  of submitted limit orders   $I_L(l)$   can be measured  from historical data. We used  the futures contract  on  index RTS and represented the rate of submitting limit orders in Figure   2 on the logarithmic scale. The vertical axis represents the number of contracts per second on both buy and sell side and the horizontal axis represents the price level.   

\begin{figure}[htp] 
\centering{
\includegraphics[scale=0.72]{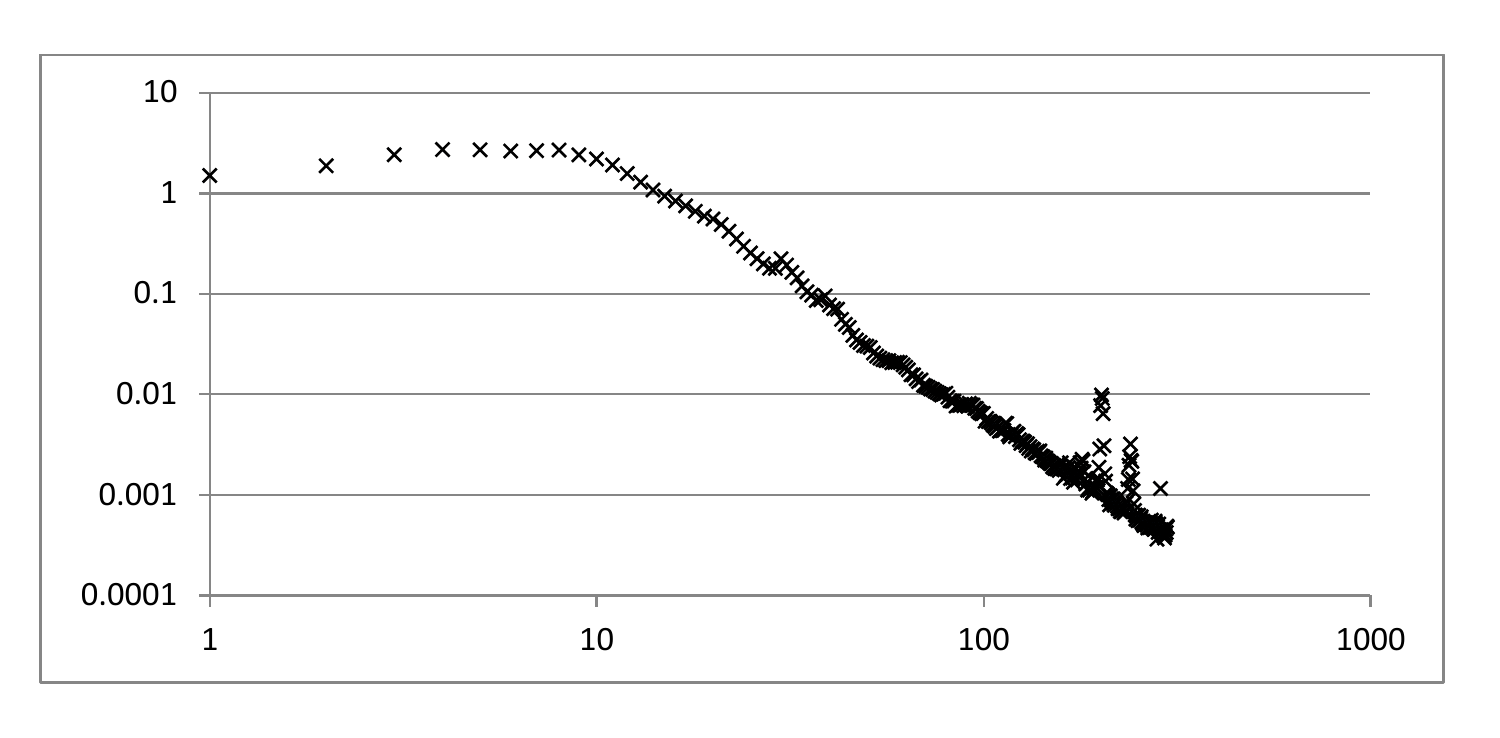}}
\caption{The rate  $I_L(l)$. }
\end{figure}

\noindent
Starting from level ten the order  submitting rate follows the power law $I_L(l)\sim l^{-\mu}$, $l > 10$.  
For RTS futures $\mu \approx 2.5$.

Similarly the rate of canceling  limit orders $I_C(l)$  on the logarithmic scale is presented in Figure 3. For level ten and higher the rate of canceling orders 
follows the   power law  $I_C(l)\sim l^{-\mu}$ with $\mu \approx 2.5$. 

\begin{figure}[htp] 
\centering{
\includegraphics[scale=0.82]{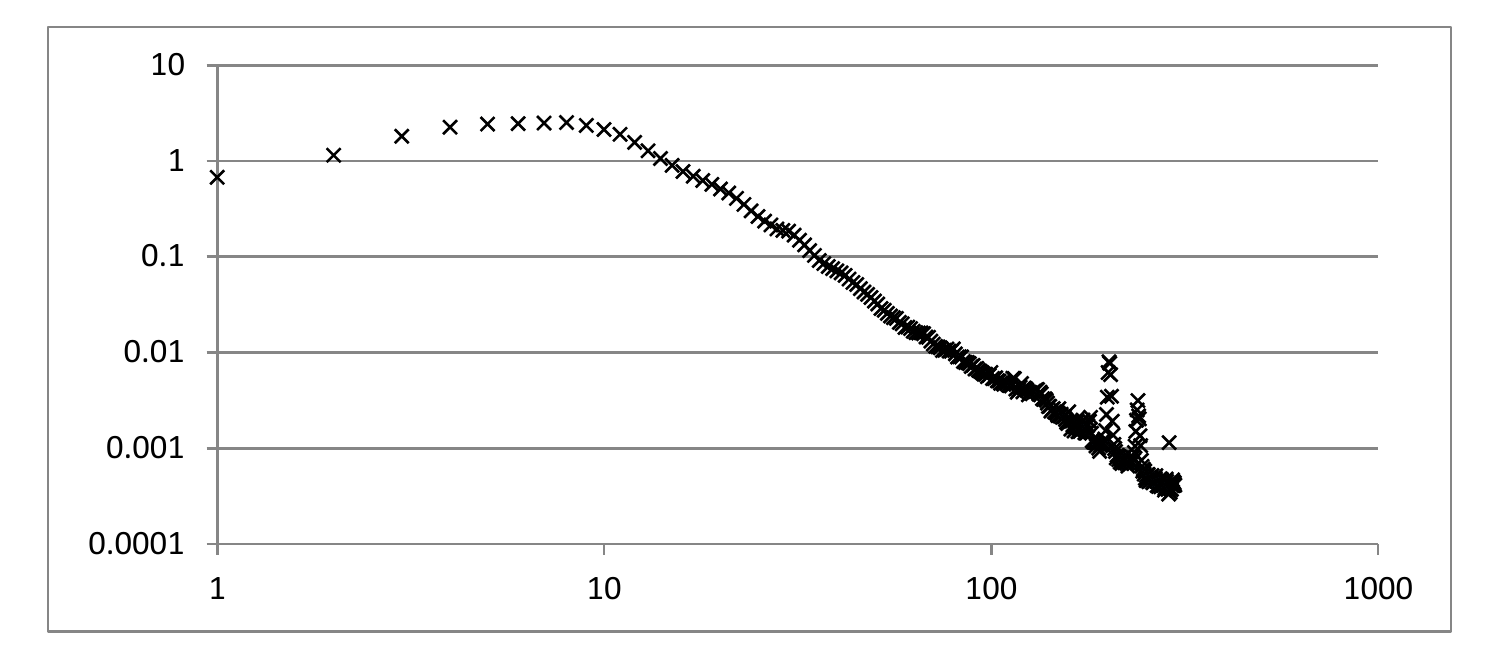}}
\caption{The rate  $I_C(l)$. }
\end{figure}

\noindent
Moreover,    $I_C(l)=I_L(l)$,  for $l>10$. 
At the same time,   $I_C(l)< I_L(l)$ for $l<10$  as shown in Figure 4
\begin{figure}[htp] 
\centering{
\includegraphics[scale=0.82]{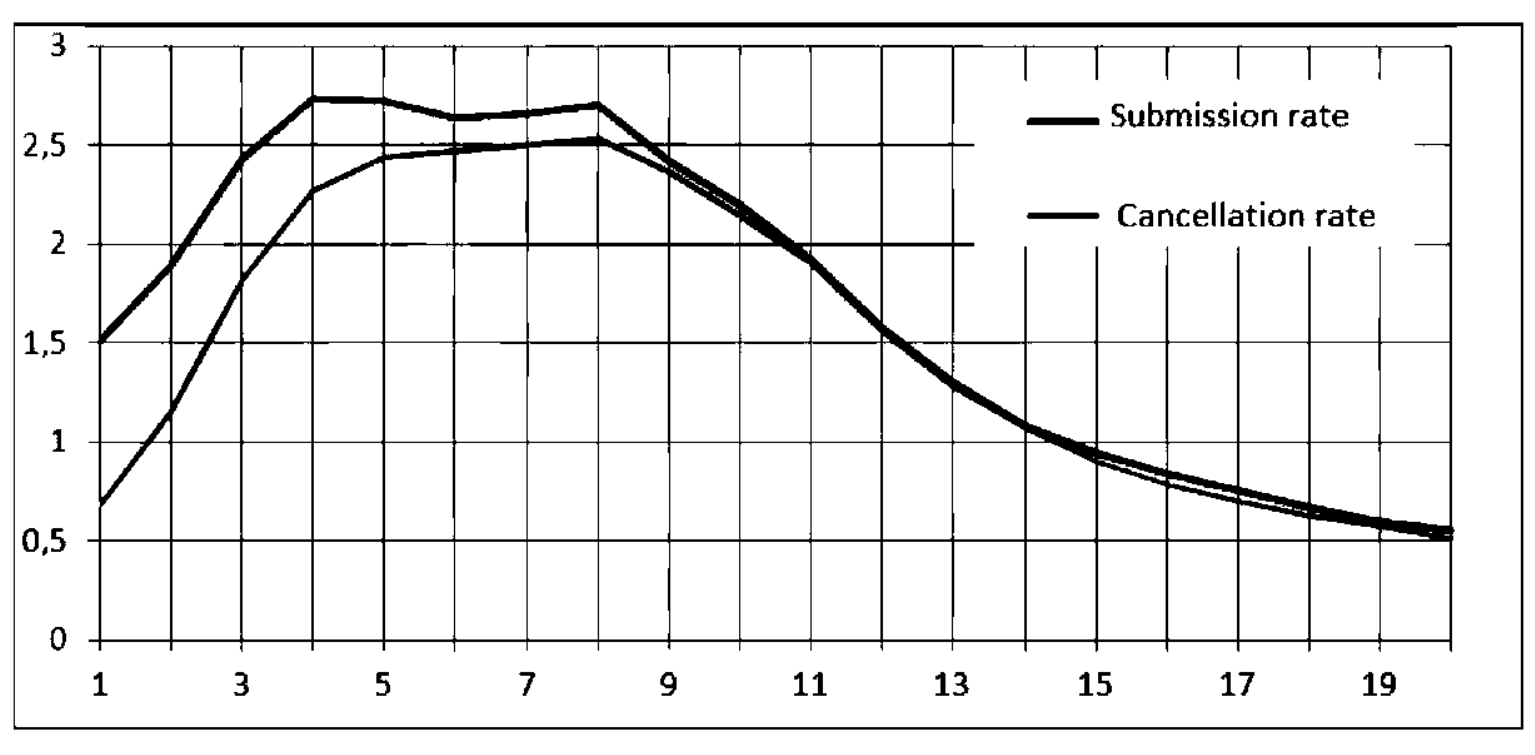}}
\caption{The smoothed curves for rates $I_C(l)$ and $I_L(l).$ }
\end{figure}

\subsection{The Order Volume.} 

The  empirical distribution  of market orders  volumes  $\hat{p}_M(\nu)$  is shown in Figure 5.  The empirical frequency is depicted on the vertical axis and 
the volume on the horizontal axis.  This  distribution can be approximated by the power law  $ \hat{p}_M(\nu)\sim \nu^{-\gamma}, \gamma\approx 2.5$. 
  
\begin{figure}[htp] 
\centering{
\includegraphics[scale=0.82]{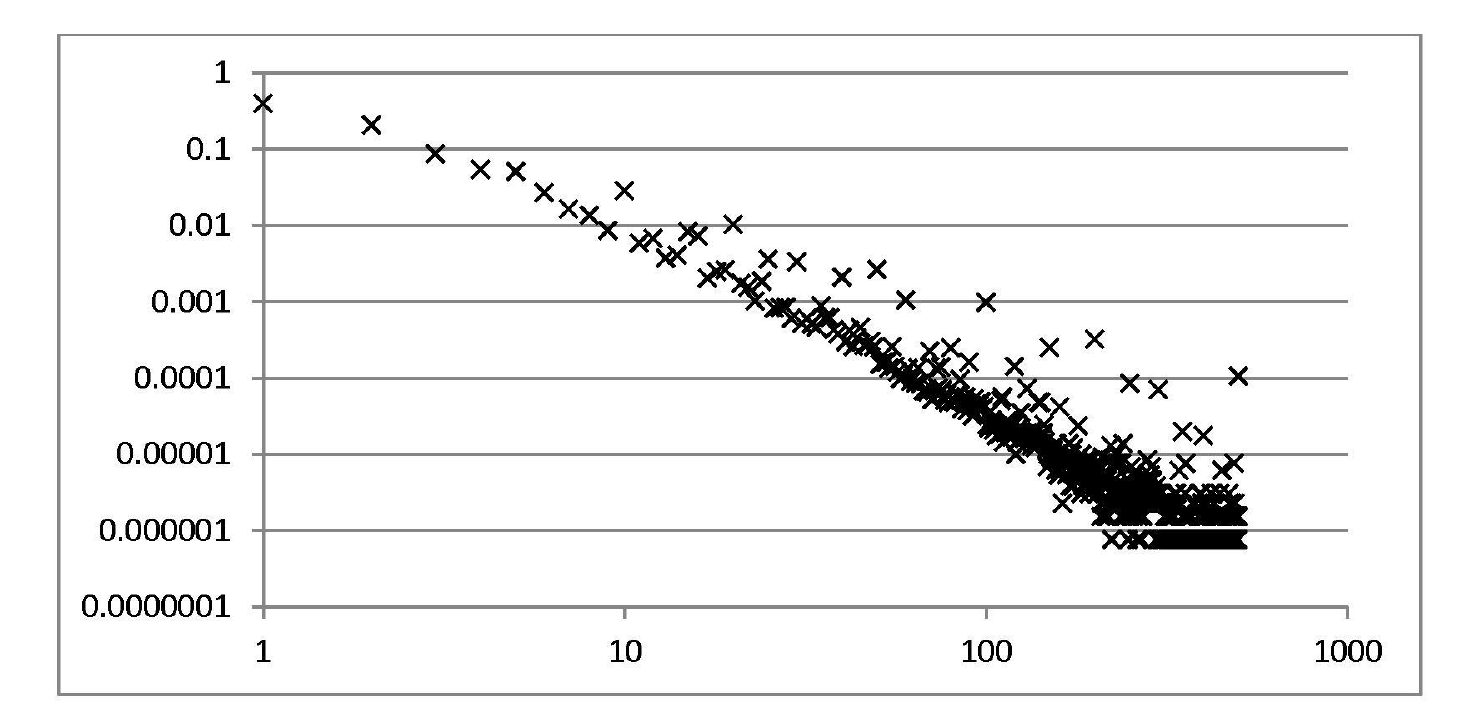}}
\caption{The empirical distribution  $\hat{p}_M(\nu)$  of volume of market orders.}
\end{figure}

\noi
The distribution  of limit order volume   $\hat{p}_L(\nu)$  is more complicated  and given in Figure 6. The volume of orders  have tendency to be multiple of 10. 
If one excludes   orders with the volume multiple of 10 then $\hat{p}_L(\nu)\sim \nu^{-\gamma}, \gamma\approx 2.8$. For orders multiple of  10 the distribution is the same with $ \gamma\approx 2.5 $. For orders multiple of 100 the law is also the same but  $\gamma\approx 2.0$. 

\begin{figure}[htp] 
\centering{
\includegraphics[scale=0.82]{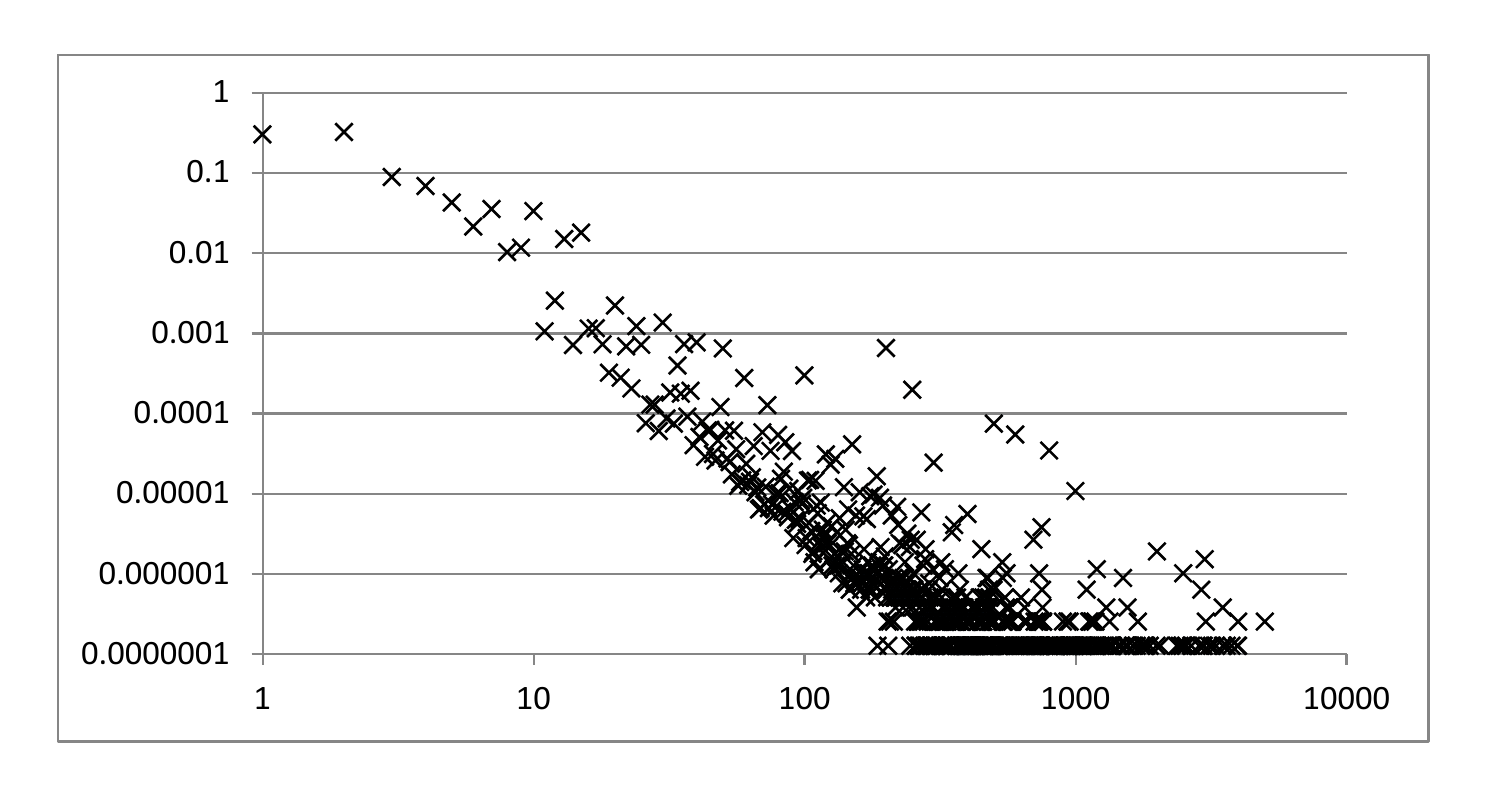}}
\caption{The empirical distribution  $\hat{p}_L(\nu)$ of volume of limit orders. }
\end{figure}

\subsection{The Book Profile.}  The book profile was determined by averaging the volume at  particular level  counted from the mid price
$$
m=\frac{a_0+b_0}{2}. 
$$
\noindent
The averaged book profile for the first 20  levels is given in Figure 7

\begin{figure}[htp] 
\centering{
\includegraphics[scale=0.82]{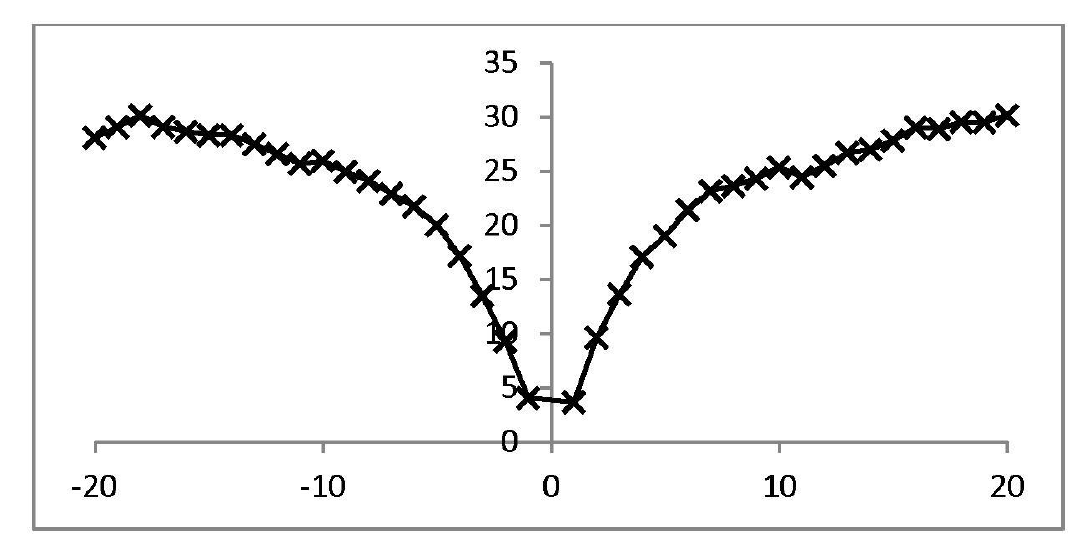}}
\caption{The book for the first 20 levels. }
\end{figure}
\noindent
The averaged book profile for the first 1000 levels is given in Figure 8 

\begin{figure}[htp] 
\centering{
\includegraphics[scale=0.82]{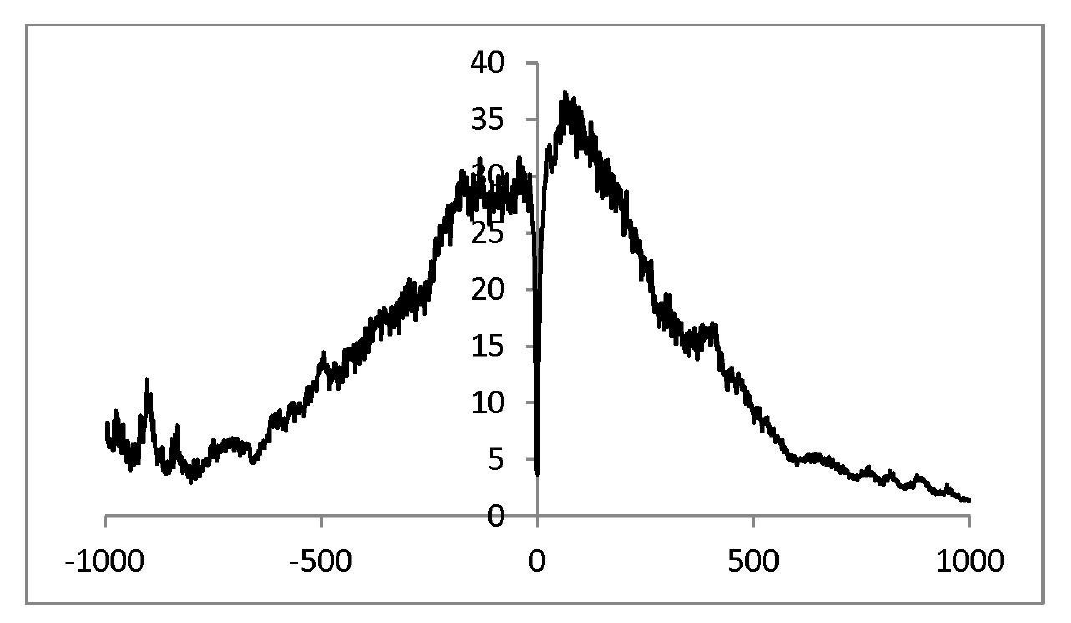}}
\caption{The book for the first 1000 levels. }
\end{figure}
\noindent
One can look at the time dynamics of the total order volume at  first one hundred levels on the sell and buy side. These are exactly the quantities 
$s(100)$ and $d(100)$ defined above. The volume is   measured for each second. 
The results are presented in Figure 9. 
\begin{figure}[htp] 
\centering{
\includegraphics[scale=0.82]{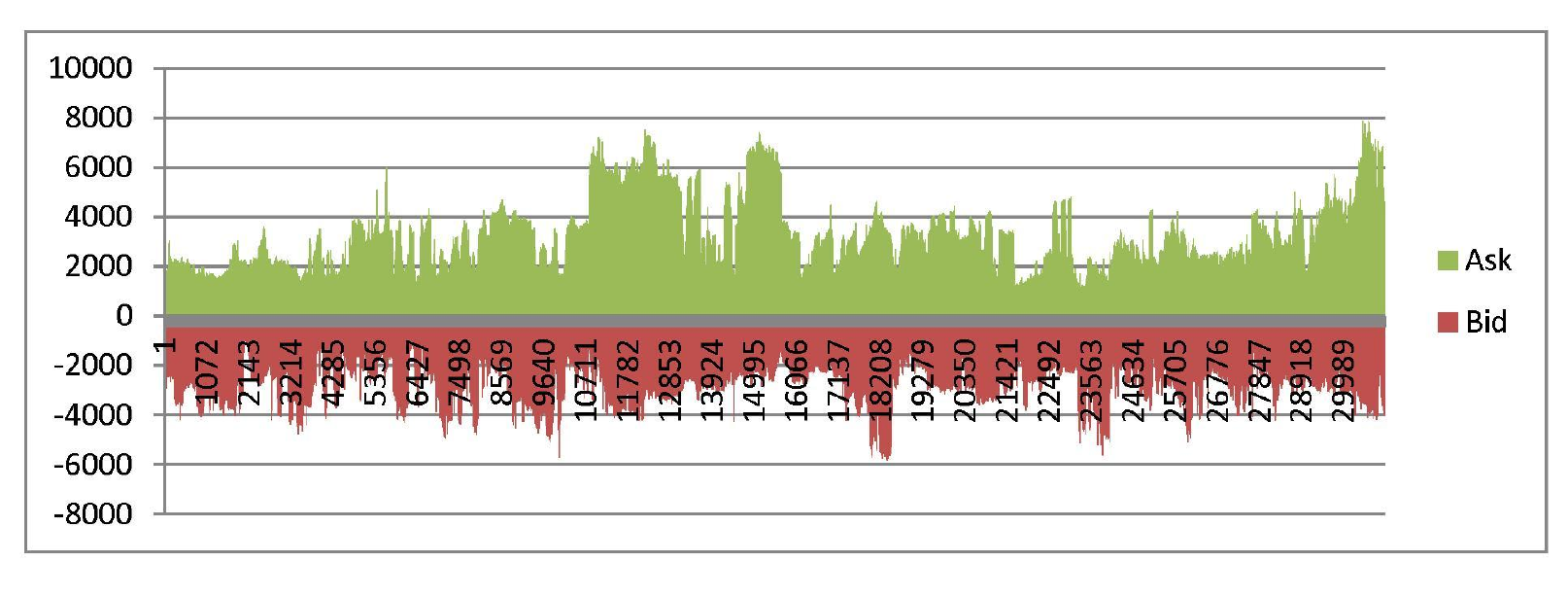}}
\caption{The  time dynamics of the aggregated volumes $s(100)$ and $d(100)$. }
\end{figure}

\subsection{The Time Between Orders.} 
The empirical distribution  of time between market orders can   measured (in seconds)  and is given in Figure 10.   

\begin{figure}[htp] 
\centering{
\includegraphics[scale=0.82]{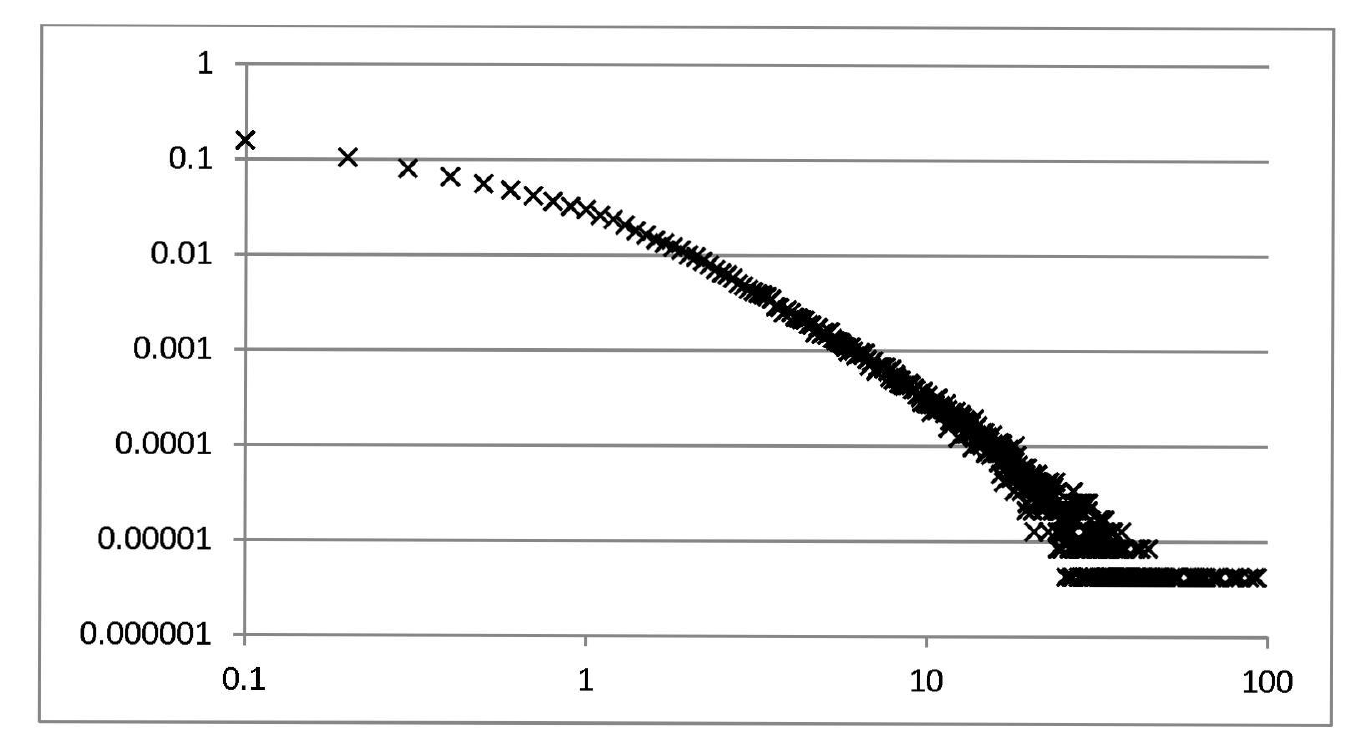}}
\caption{The empirical   distribution of time between market orders. }
\end{figure}

The empirical distribution of time between limit orders can also be measured (in seconds) and is  presented in Figure 11
\begin{figure}[htp] 
\centering{
\includegraphics[scale=0.82]{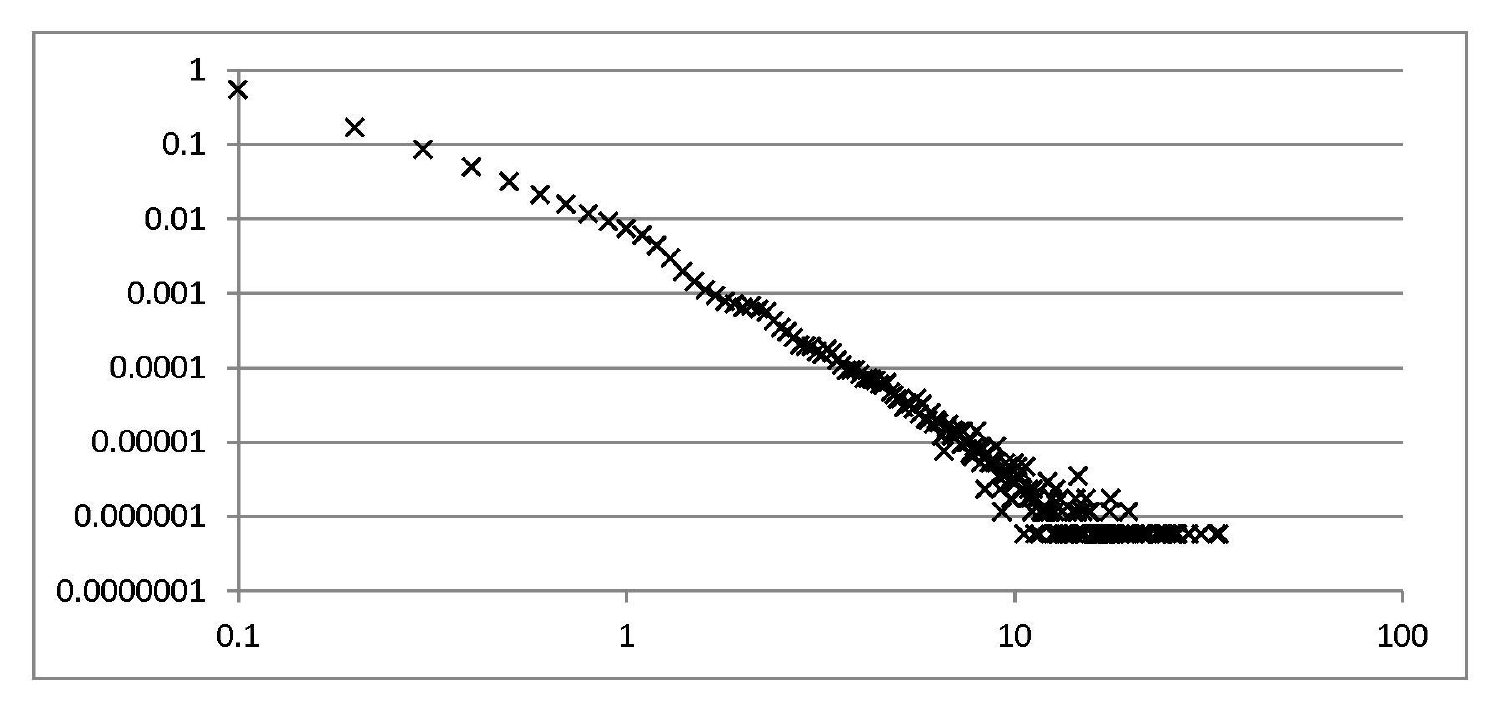}}
\caption{The empirical distribution of  time between limit orders. }
\end{figure}

\newpage

\section{The Poissonian Multi--Agent Model.}  

\subsection{Market Participants}

\noi
In accordance with a mechanism of the double auction there are six type of events that can occur in the order book

\begin{itemize}
  \item Liquidity provider submits  buy  limit order
  \item Liquidity provider submits sell limit  order
  \item Liquidity taker  submits market  buy  order
  \item Liquidity taker  submits market sell  order 
  \item Liquidity taker  cancels buy limit order
  \item Liquidity taker cancels sell limit order
\end{itemize}

These events can be produced  by six   agents or market participants.  One group   are agents providing liquidity to the book and another group 
are agents taking liquidity from the book.

One group are liquidity providers which  differ from each other  by a direction of limit  orders.  Providers of sell liquidity submit sell limit orders to the book and  
providers of buy liquidity submit buy limit orders to the book. 

Another group  are liquidity takers. 
Liquidity takers submit market 
orders to the book and also  differ from each other  by a  direction. For example,  
liquidity takers  of buy orders send  market   sell orders to the book. Similarly liquidity takers of sell 
orders send buy  market buy orders.   Another type  of liquidity takers   can   cancel active buy or sell limit orders.

Everywhere below we assume that the events in our model  form  a Poisson flow. Namely the time  $\tau$  between two consecutive events is exponentially distributed 
with  the distribution $Prob\, \{ \tau \geq t\}= \exp (-I t),$ where the parameter $I   >0$ is called the rate.

Parameters for providers and takers of $buy$  liquidity we denote with the lower subscript $bid$ and parameters of  takers and providers of $sell$ liquidity  are denoted with the subscript $ask$. The superscript specifies the type and stands for the following 
\begin{itemize}
\item  $L$  -- limit order,
\item  $M$  -- market order,
\item  $C$  -- cancellation of an active limit order in the book. 
\end{itemize}

Every group of market participants acts with the Poisson rate $I^{type}_{side}$ where the subscript stands for the direction and the superscript for the type of action. The total rate of all events in the exchange is given by the formula
$$
I= I_{ask}^{L} + I_{bid}^{L} +I_{ask}^{C} + I_{bid}^{C} +I_{ask}^{M} + I_{bid}^{M}; 
$$
where 

 $I_{bid}^{L}$ -- the rate of submitting limit buy orders;

 $I_{ask}^{L}$  -- the rate of submitting limit sell orders;
 
 $I_{ask}^{M} $  -- the rate of submitting market buy orders;
 
 $  I_{bid}^{M}$ --  the rate of submitting market sell orders;
  
 $I_{ask}^{C} $ -- the rate of submitting cancellation request for limit sell orders;
  
 $ I_{bid}^{C} $  -- the rate of submitting cancellation request for limit buy orders.

\noi
On each step of simulation only one of these six events occurs. The time between two consecutive events is exponentially distributed with the rate $I$. The  probability 
of  an event is given by the formula 
$$
\frac{I^{type}_{side}}{I}. 
$$
For example,   the probability of cancellation of some buy limit  (bid)  order is 
$$
\frac{I^{C}_{bid}}{I}. 
$$

\subsection{Liquidity providers.}
Liquidity providers submit limit orders buy or sell of volume $\nu$ at some  price level $l$. Price level $l$ takes also integer values $1,2,\cdots,K;$ with the probability $q^L(l).$ We assume also that maximal price level $K=1000$. The distribution function $q^L(l)$ is determined by the empirical rate $I_L(l)$. The volume $\nu$  of  an order   takes integer values $1,2,\cdots, V^L;$   with probability $p^L(\nu).$  We assume that  maximal volume $V^{L}=1000.$  The distribution $p^L(\nu)$  is 
modeled upon the empirical distribution $\hat{p}^L(\nu)$. 
The distribution functions for the volume and the price level are the same 
for providers of sell and buy orders.

The limit orders can be executed partially, meaning that if they are bigger then the size of a market order then just some part of them can be filled.  

The rate of submitting liquidity {\it i.e.} limit orders in the book
$$
V_{in}=S^{L} (I_{ask}^{L} + I_{bid}^{L} ),   
$$
where
$$
S^{L}= \sum_{\nu=1}^{V^L} \nu p^L(\nu). 
$$

\subsection{Liquidity takers.} Liquidity takers submit either market orders or just simply cancel existent limit orders in the book. 
Market orders have a random volume $\nu$ which takes values $1,2,\cdots, V^M;$ with probability $p^M(\nu)$ which is modeled upon the empirical distribution  $\hat{p}^M(\nu)$.  The maximal volume $V^M=100$.

\subsection{Conditions of equilibrium.} 
Cancellation of limit orders happens with an equal probability for all active limit orders buy or sell. Let us define 
$$
S^{C}= \sum_{\nu=1}^{V^L} \nu p^C(\nu),
$$
where $p^C(\nu)$ is the distribution function  of the volume of canceled limit orders. 
Active limit orders in the book are subjected to the flow of market orders. 
Market orders  can take limit orders completely or just make the  size of  limit orders smaller  than when they were actually submitted. Therefore,  
$$
S^C < S^{L}. 
$$

The rate of liquidity consumption 
$$
V_{out}= S^M (I_{ask}^{M} + I_{bid}^{M } ) + S^C (I_{ask}^{C} + I_{bid}^{C} ),  
$$
where 
$$
S^{M}= \sum_{\nu=1}^{V^M} \nu p^M(\nu).
$$

The quantities $s$ and $d$ determine instant liquidity in the book. The rate of change of instant liquidity is given by 
$$
\Delta s=I_{ask}^L  S^{L} - I_{ask}^{M}  S^{M}- I_{ask}^C S^{C}, 
$$
and
$$
\Delta d=I_{bid}^L  S^{L} - I_{bid}^{M}  S^{M} - I_{bid}^C S^{C}.
$$
Obviously  in the  stationary regime $\Delta s=\Delta d=0$  and   the following identities hold
$$
S^{L} I_{ask}^L= S^M I_{ask}^M+ S^C I_{ask}^C, 
$$
$$
S^{L} I_{bid}^L= S^M I_{bid}^M+ S^C I_{bid}^C.
$$
These imply $V_{in}=V_{out}$. 

When market orders are not present  $I_{bid}^M=I_{ask}^M=0$  we have $S^C=S^L$ and therefore
\bey
I_{bid}^L&=& I_{bid}^C,\\
I_{ask}^L&=& I_{ask}^C. 
\eey

Let us also define aggregated supply   of sell orders 
$$
S=I_{ask}^L  S^{L}+ I_{bid}^{M}  S^{M}- I_{ask}^C S^{C}, 
$$
and buy orders 
$$
D=I_{bid}^L  S^{L}  + I_{ask}^{M}  S^{M}  - I_{bid}^C   S^{C}.
$$
The state of  equilibrium is determined by  two conditions  
\begin{itemize}
  \item The balance of instantaneous liquidity.
  \item The balance  of rate of change of instant liquidity. 
\end{itemize}  
These conditions imply $s=d  >0$ and $\Delta s=\Delta d$  which is equivalent to $S=D$. 
We are going to study price dynamics in terms $s,\,d,\,S,\,D$. 

\section{Results of Simulation.}

We would like to note that sometime during simulation there are no limit orders in the  book  on   sell or buy side. 
In other words,  due to randomness liquidity in the book can drop to zero, meaning  $s=0$ or $d=0$. In such case  when 
market order arrives it will be no limit orders  in the book to match market order. 
In order to avoid this  we impose the following conditions 
$$
s > s_{min},  \qquad\qquad d >  d_{min}, 
$$ 
where $s_{min} > V^M$   and $d_{min}  >   V^M$.  
Once any of these conditions has been  violated   we need to stop the flow of market orders and also stop cancellations 
$$
I_{bid}^{M}=I_{bid}^C=0,\qquad \text{if} \quad d  < d_{min}, 
$$
or  
$$
I_{ask}^{M}=I_{ask}^C=0,\qquad \text{if} \quad s  < s_{min}.
$$

Another problem in running simulations is an unlimited growth of a number of limit orders in the book;   in other words instant liquidity can not grow indefinitely. 
We arrange parameters (the rates $I_{side}^{type}$)   such that aggregated rate    of liquidity supply is less than  aggregated  rate   of liquidity consumption.   
This implies that the rates  have to be such that the following conditions hold
$$
\Delta  d < 0, \qquad \text{for} \quad d  > d_{min}, 
$$
and
$$
\Delta  s < 0, \qquad \text{for} \quad s  > s_{min}. 
$$

\subsection{The Profile of the Book  and the Spread.}

Let us define profile of the book as the  state of all queues at a particular moment of time. Average profile is computed by averaging instantaneous 
profiles for each second on a particular time interval. 

The response of  the book profile to the flow of market orders  can be easily understood. When the market orders are absent $I_{bid}^M=0$  and 
$I_{ask}^M=0$ all existent  orders are canceled without exception  and $S_L=S_C$. This implies that 
$$
I_{ask}^L   = I_{ask}^C,\qquad\qquad \qquad  I_{bid}^L   = I_{bid}^C. 
$$
Since in our model the size of limit order is independent from the price and direction of the trade then
the the book is filled uniformly with the limit orders  as it is shown in Figure 12.  
\begin{figure}[htp] 
\centering{
\includegraphics[scale=0.62]{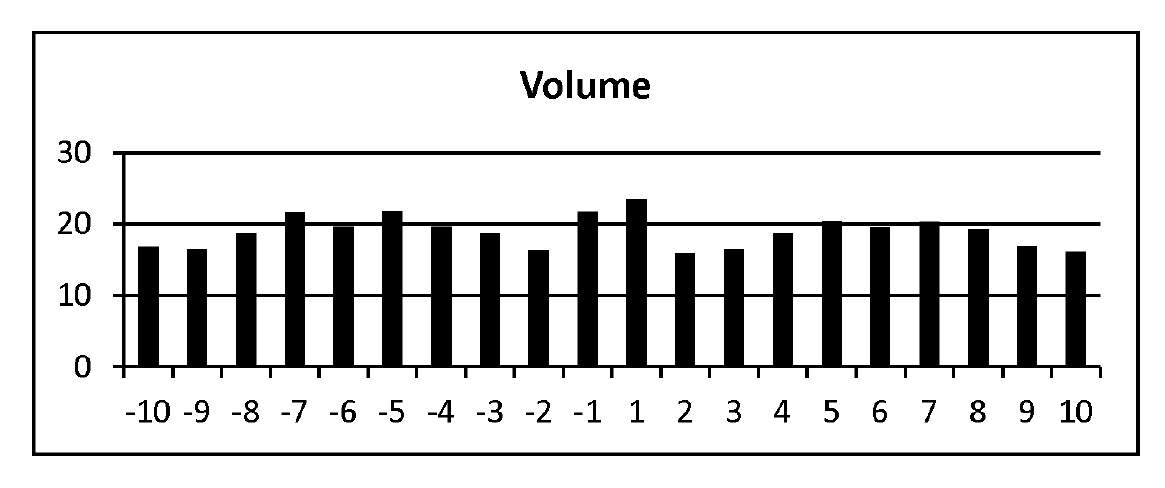}}
\caption{The Book profile  without market orders. }
\end{figure}
Consider now  the spread 
$$
\delta=\frac{p_{ask}-p_{bid}}{s},
$$
and let us study how its depends on the size of a market order.

When the  market orders submission rate is small
$$
\frac{I_{ask}^M}{I}< 0.01,\quad\quad\quad\quad\quad\quad    \frac{I_{bid}^M}{I}< 0.01, 
$$
then the  book profile remains unchanged  as shown in Figure 13  

\begin{figure}[htp] 
\centering{
\includegraphics[scale=0.62]{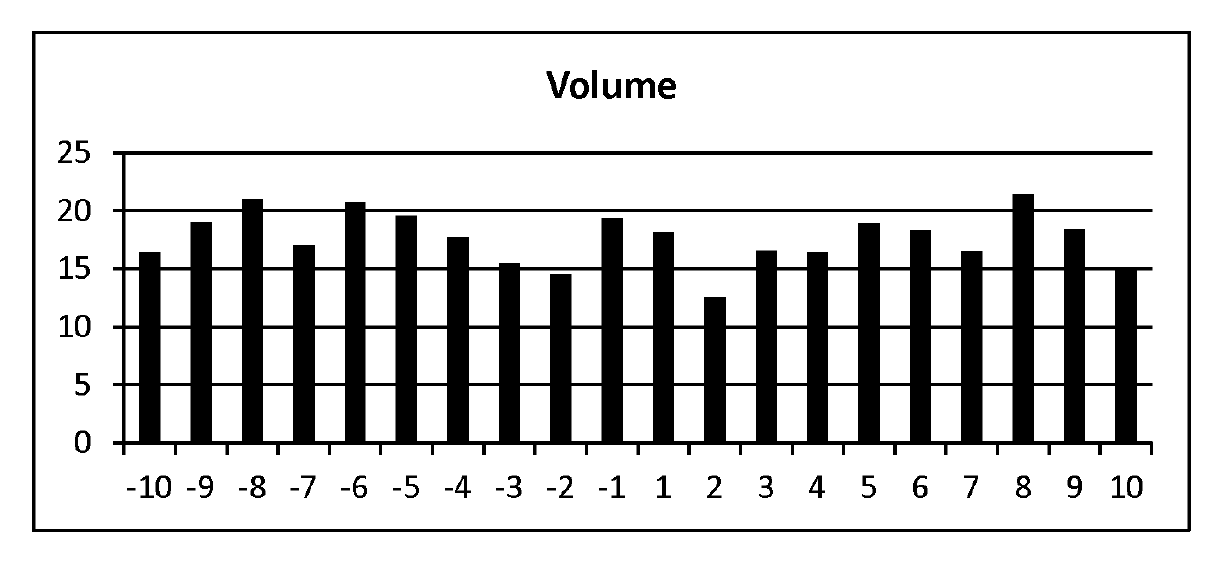}}
\caption{The Book profile  with small rate of market orders. }
\end{figure}

The spread (after the market order has been  executed)  depends  linearly 
on the size of a market order
$$
\delta \sim v^M. 
$$ 
The empirical relation between the size and a spread is depicted in Figure  14. The size of limit order is depicted on the vertical axis  and the 
spread is shown on the horizontal axis.

\begin{figure}[htp] 
\centering{
\includegraphics[scale=0.62]{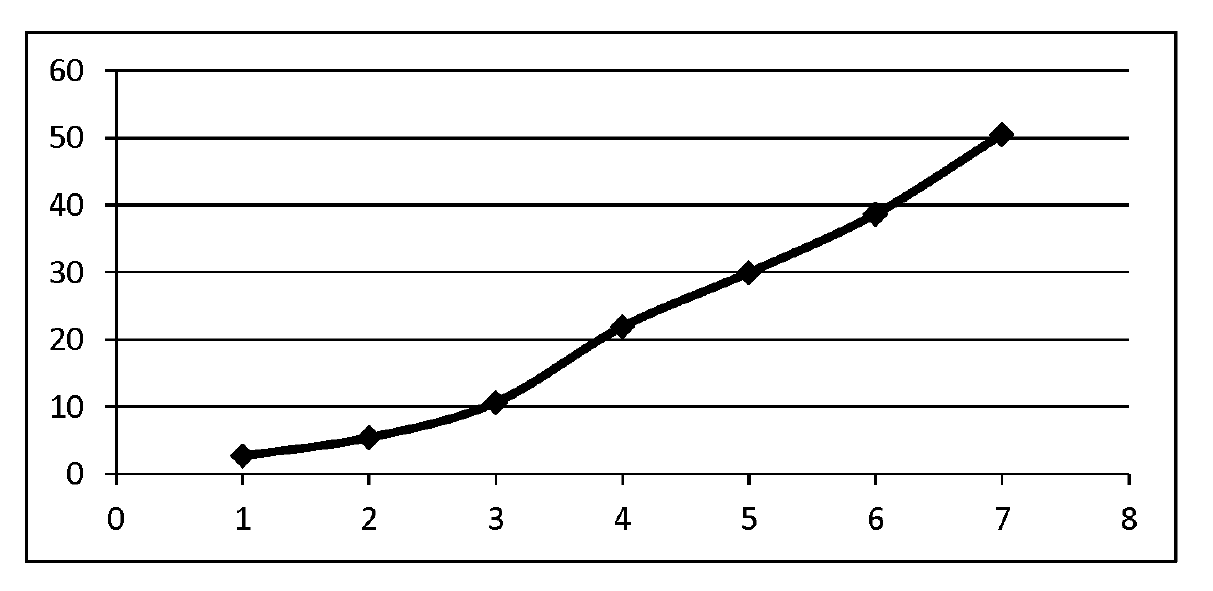}}
\caption{The size of market orders as the function of spread.}
\end{figure}

When the rate  of market orders increases 
$$
\frac{I_{ask}^M}{I}  \sim 0.1,\quad\quad\quad\quad\quad\quad    \frac{I_{bid}^M}{I}  \sim 0.1, 
$$
then the book profile changes. On the levels closest to the bid or ask the size of the book is almost linearly 
depends on a level number  as shown in Figure 15 
$$
X_{i_{bid}+l} \sim l, \quad\quad X_{i_{ask}-l} \sim l,  
$$
where $l >0$. When market order arrives it annihilates  limit orders at the level proportional to the square root of the volume and 
$$
\delta  \sim \sqrt{v^M}. 
$$

\begin{figure}[htp] 
\centering{
\includegraphics[scale=0.82]{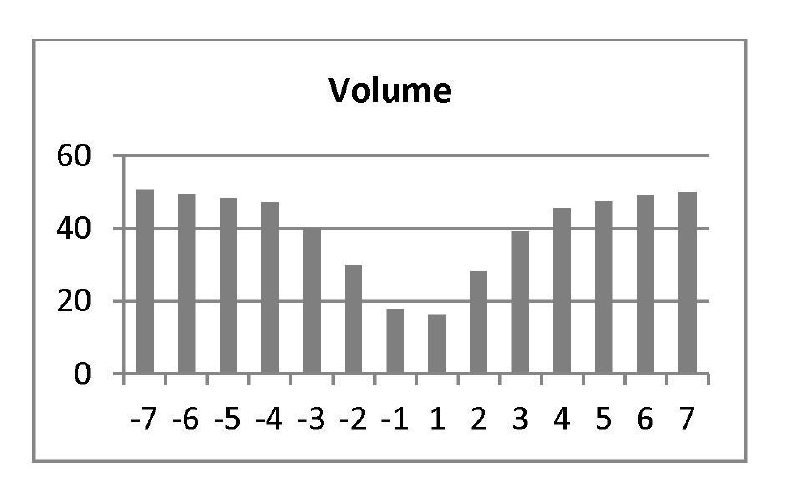}}
\caption{The book profile  with the high rate of market orders. }
\end{figure}

\subsection{The Balance of Liquidity in the Book  and in the Order Flow. } 

We define  the parameter $I=179$ {\it events/second}. All other parameters in the balanced case are given in the table 

\begin{figure}[htp] 
\centering{
\includegraphics[scale=0.82]{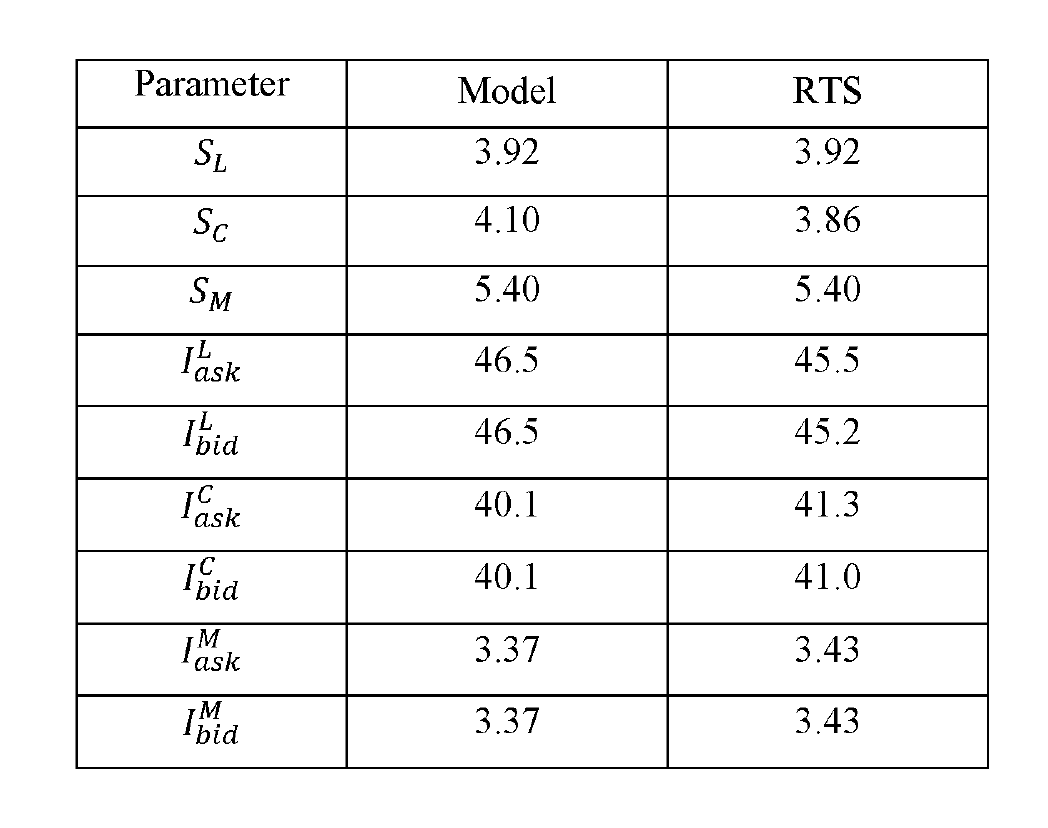}}
\caption{Parameters in the balanced case.}
\end{figure} 

The evolution of the price in our model is given below
 \begin{figure}[htp] 
\centering{
\includegraphics[scale=0.62]{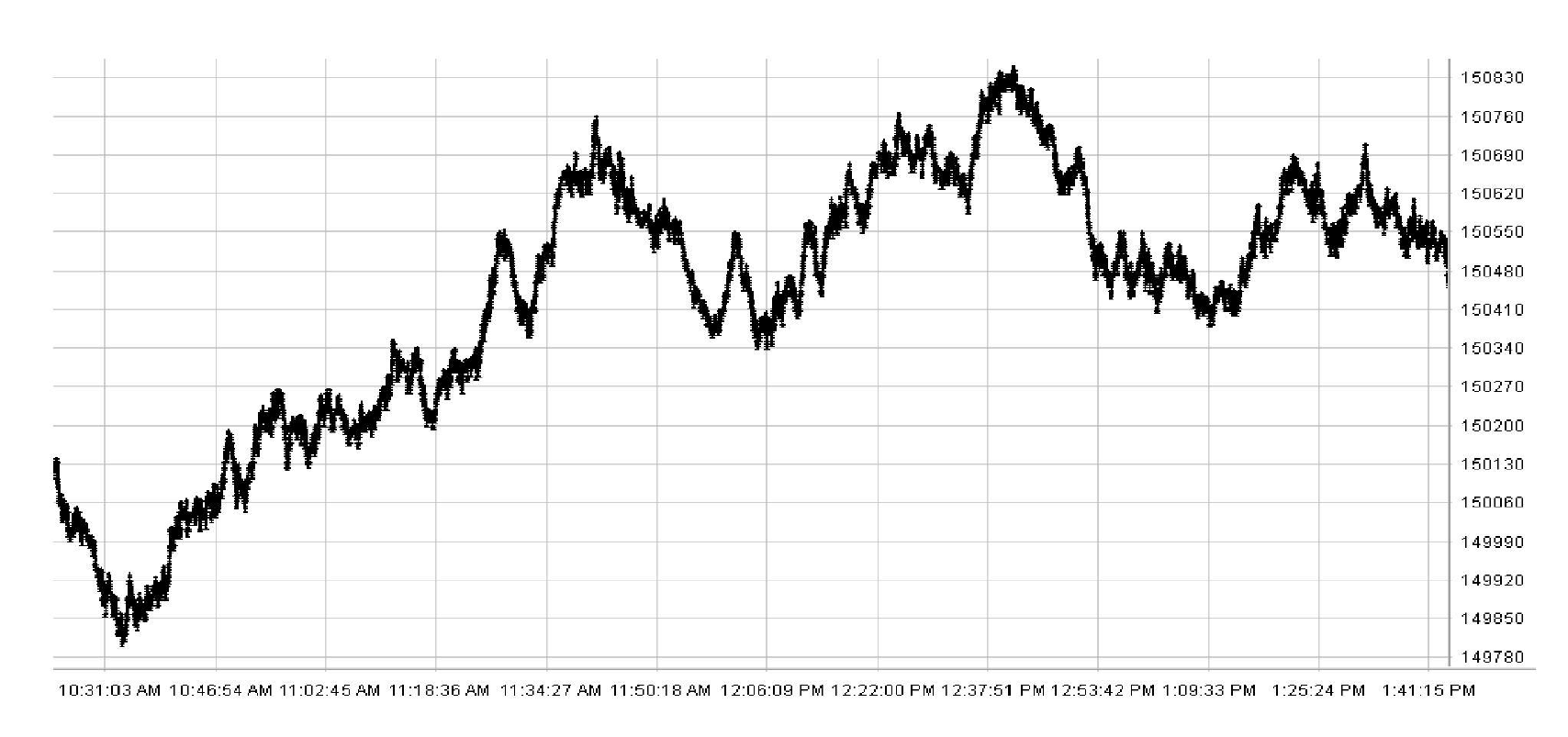}}
\caption{Price in the balanced case.}
\end{figure} 
\noindent

\noindent
Apparently the price does not exhibit  any preferred direction. 
We refer to \cite{G} for details of this simulation. 

\subsection{The Disbalance of Liquidity in the Book. }  

By adjusting $s_{min}$ and $d_{min}$ one can  model price movements. 
We assume that the all other rates on the sell and buy side are equal  
$$
I_{bid}^M=I_{ask}^M,\qquad\qquad I_{bid}^C= I_{ask}^C,\qquad\qquad I_{bid}^L= I_{ask}^L. 
$$
If $s_{min} > d_{min}$ then  the book is  thiner on the buy side (below  the  the price) and this leads to price decrease. If on the opposite 
$s_{min} < d_{min}$ then  the book is  thiner on the sell  side (above   the  the price) and this leads to price increase.

Indeed, the dependence of $\delta_{sell}$ on the volume of sell market order $v_M$ is getting bigger as soon as $d_{min}$ is getting smaller. Similarly, dependence 
of $\delta_{buy}$ on the volume of buy market order $v_M$ is getting bigger as soon as $s_{min}$ is getting smaller.  As a very crude approximation we can take 
$$
v_M\sim \delta_{sell} d_{min}, 
$$
and 
$$
v_M\sim \delta_{buy} s_{min}.
$$
Therefore, 
$$
\frac{\delta_{sell}}{\delta_{buy}} \sim \frac{s_{min}}{d_{min}}. 
$$
If $s_{min} < d_{min}$,  then   
$$\frac{\delta_{sell}}{\delta_{buy}}  <1$$ and the price has to increase.  If $s_{min} > d_{min}$,  then 
$$\frac{\delta_{sell}}{\delta_{buy}}  > 1$$ and  the price has to decrease.

\begin{figure}[htp] 
\centering{
\includegraphics[scale=0.52]{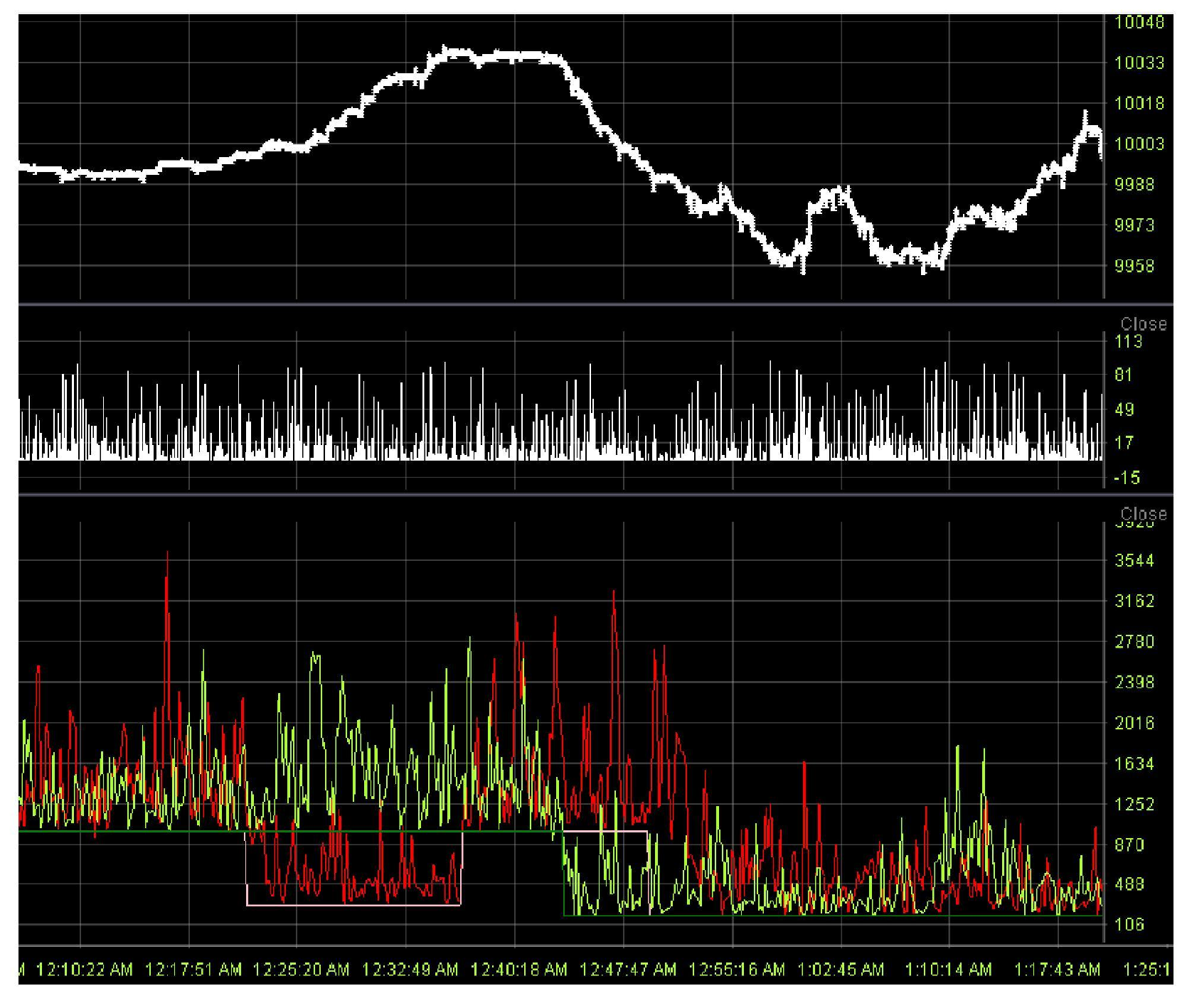}}
\caption{Thinning of the  book on the buy or sell side. }
\end{figure}

This is illustrated in Figure 18. The first graph is  a price  and the second graph which is  the vertical column is the volume. The third graph  is the graph for 
instantaneous liquidity   $s$ and $d$. The white and green lines are the graphs $s_{min}$ and $d_{min}$. Depending on the relation between $s_{min}$ and $d_{min}$ one can observe increase or decrease of the price.

\subsection{The Disbalance of Sell and Buy orders in the Order Flow.}

Such disbalance occurs when
$
I_{bid}^M\neq I_{ask}^M.
$ 
At the same time the condition $\Delta s =\Delta d=0$  holds. Figure 19 shows monotonous increase of the price.  
The first graph represent the price 
and the second two red and yellow lines are $s(5)$ and $d(5)$.

\begin{figure}[htp] 
\centering{
\includegraphics[scale=0.62]{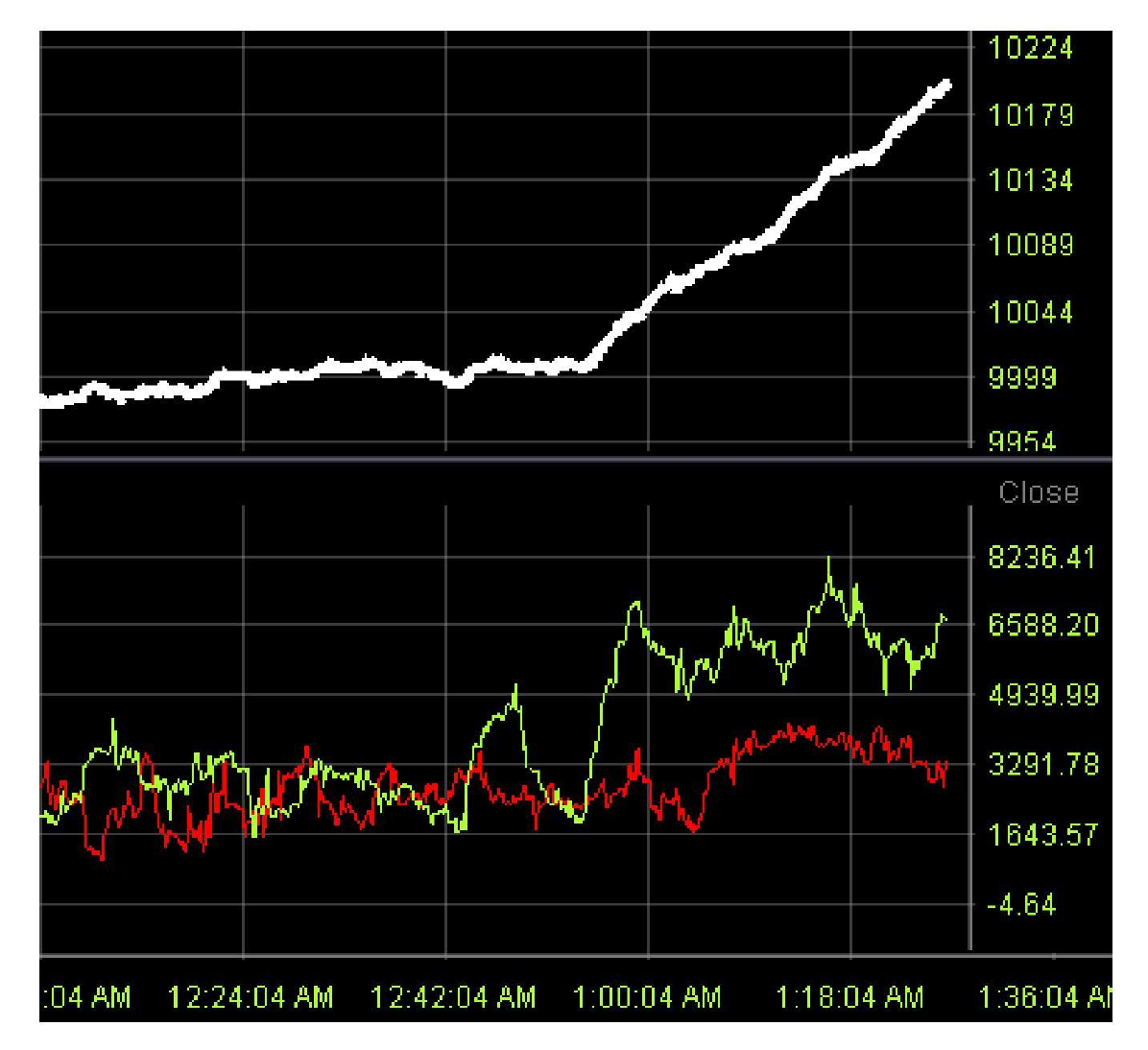}}
\caption{The upward trend. }
\end{figure}

Figure 20  also shows   monotonous increase of the price but instead of $s(5)$ and $d(5)$   it  has graphs of $s$ and $d$.  

\begin{figure}[htp] 
\centering{
\includegraphics[scale=0.82]{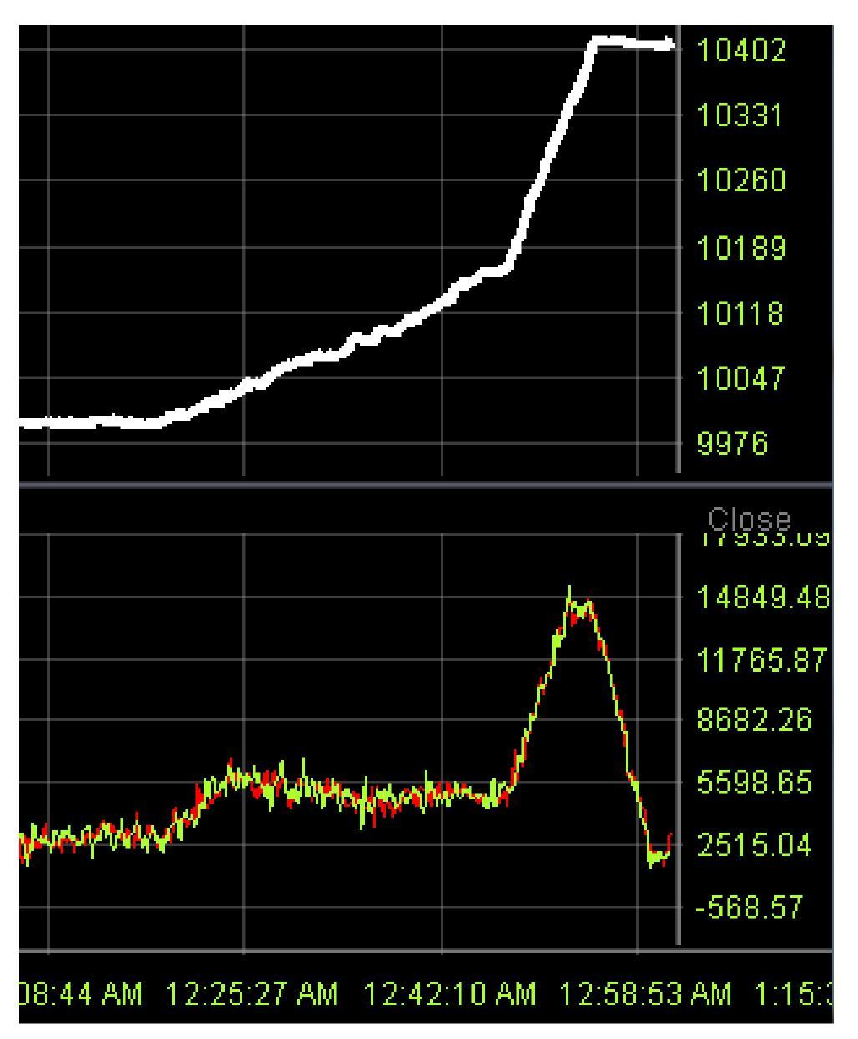}}
\caption{The upward trend. }
\end{figure}

\vskip .2in
\noindent 
Kirill Vaninsky
\newline
Department of Mathematics
\newline
Michigan State University
\newline
East Lansing, MI 48824
\newline
USA
\noindent
\newline
vaninsky@math.msu.edu

\vskip .2in
\noindent
Alexander Glekin
\newline
Institute of System Analysis 
\newline
RAN 
\newline 
Prospect 60 Oktyabria
\newline
Moscow 117312
\newline
Russia
\noindent
\newline
AleksanderGlekin@gmail.com

\vskip .2in
\noindent
Alexander Lykov 
\newline
Faculty of Mathematics and Mechanics 
\newline
Moscow State University 
\newline 
Vorobjevy Gory
\newline
Moscow
\newline
Russia
\noindent
\newline
alexlyk314@gmail.com

\newpage

\end{document}

%% file: definitions.tex
\usepackage{graphicx}
\usepackage{amsmath}

\newcommand{\noi}{\noindent}

\def\[{\left[}
\def\]{\right]}
\def\({\left(}
\def\){\right)}

\newcommand{\eeq}{\end{equation}}
\newcommand{\beq}{\begin{equation}}
\newcommand{\bay}{\begin{eqnarray}}
\newcommand{\ey}{\end{eqnarray}}
\newcommand{\bey}{\begin{eqnarray*}}
\newcommand{\eey}{\end{eqnarray*}}

